\documentclass[twocolumn,showpacs,preprintnumbers,amsmath,amssymb]{revtex4}

\usepackage{graphicx}
\usepackage{dcolumn}
\usepackage{bm}

\newcommand{\sign}{\operatorname{sign}}
 
\begin{document}


\title{Diluted 3d-Random Field Ising Model at zero temperature with
  metastable dynamics}

\author{Xavier Illa} \email{xit@ecm.ub.es} \author{Eduard Vives}
\email{eduard@ecm.ub.es}
\affiliation{%
  Departament d'Estructura i Constituents de la Mat\`eria,
  Universitat de Barcelona \\
  Mart\'{\i} i Franqu\`es 1, Facultat de F\'{\i}sica, 08028 Barcelona,
  Catalonia}

\date{\today}

\begin{abstract}
  We study the influence of vacancy concentration on the behaviour of
  the three dimensional Random Field Ising model with metastable
  dynamics.  We focus our analysis on the number of spanning
  avalanches which allows for a clean determination of the critical
  line where the hysteresis loops change from continuous to
  discontinuous. By a detailed finite size scaling analysis we
  determine the phase diagram and estimate numerically the critical
  exponents along the whole critical line. Finally we discuss the
  origin of the curvature of the critical line at high vacancy
  concentration.
\end{abstract}



\maketitle

\section{Introduction}
\label{Intro}

Externally driven systems at low enough temperature often display rate
independent hysteresis. This out-of-equilibrium phenomenon occurs
because intrinsic disorder creates multiple energy barriers that the
system cannot overcome due to the very weak thermal fluctuations.

The study of zero temperature models with metastable dynamics has been
very succesful for understanding rate independent hysteresis. A
prototype case is the Random Field Ising Model (RFIM) with single
spin-flip relaxation dynamics \cite{Sethna1993,Sethna2005}. 
Although the model is
formulated in terms of magnetic variables (external field $H$ and
magnetization $m$) it can be applied to the study of many phenomena
associated to low temperature first-order phase transitions in
disordered systems, e.g. martensitic transformations \cite{Ortin2005},
fluid adsorption in porous solids \cite{Detcheverry2003},
ferroelectrics \cite{Tadic2002}, etc.

Disorder is a intriguing concept: in the RFIM it is introduced via
independent and quenched random fields on each lattice site, gaussian
distributed with zero mean and standard deviation $\sigma$.  In real
materials, disorder is much more complicated and includes features at
all length scales: vacancies, interstitials, composition fluctuations,
dislocations, strain fields, grain boundaries, sample surfaces, edges
and corners, etc.  Thus it is interesting to add to the RFIM other
sources of disorder in order to see how the non-equilibrium behaviour
is modified.

The goal of this paper is to study the diluted RFIM at $T=0$ with
metastable dynamics and analyze the consequence of introducing a
concentration $c$ of quenched vacancies. The interplay between the two
kinds of disorder (random fields and vacancies) will be at the origin
of the properties of the $\sigma-c$ phase diagram.

One of the striking results concerning the RFIM with metastable
dynamics, as already pointed out in the seminal paper of Sethna et
al.\cite{Sethna1993}, is the occurence of a critical point when the
amount of disorder $\sigma$ is increased.  The $m$ vs. $H$ hysteresis
loops change from discontinuous (like in a ferromagnet) when
$\sigma<\sigma_c$ to continuous (like in a spin-glass) when
$\sigma>\sigma_c$.  This result was demonstrated using mean-field
analysis and numerical simulations in 3d systems. This problem was
also studied within the Renormalization Group formalism
\cite{Dahmen1993, Dahmen1996}.  Moreover, many properties of the
critical point have been also studied analytically on Bethe lattices
\cite{Dhar1997,Sabhapandit2000,Shukla2001,Illa2005,Illa2006}.
Experimental evidence for the occurence of such a critical point 
has been found in different magnetic systems \cite{Berger2000,Marcos2003}.

Another interesting result of the RFIM with metastable dynamics is
that it reproduces the experimental observation that the $m(H)$
trajectories of such athermal systems are discontinuous at small
scales. The evolution proceeds by avalanches from a mestastable state
to another. In the RFIM the avalanche size distribution becomes a
power-law at the critical point. Experimentally, scale free
distributions of avalanche properties have been found in many systems
\cite{Babcock1990,Cote1991,Lilly1993,Vives1994,Wu1995,Carrillo1998,
Puppin2000,Durin2000}. A
first attempt to study the influence of dilution in such avalanche
size distributions was done some years ago \cite{Tadic1996}. The
results of this work, however, should be considered as only
qualitative, given the fact that the studied system was
two-dimensional \footnote{For a discussion of the problems in the
  non-diluted 2d RFIM, see Ref.~\onlinecite{Perkovic1996}.}, the
analysis focused only on the avalanche distributions and the results
concerning the phase diagram were very approximate.

The order parameter that vanishes at the critical point is the size of
the macroscopic discontinuity $\Delta m$. The analysis of this
quantity from simulations is very intrincate. In finite-size systems
it is very difficult to make the distinction between a macroscopic
jump and a microscopic avalanche.  The measured order parameter only
displays reasonable finite-size scaling (FSS) properties when the
simulated systems are very large \cite{Kuntz1999}. Recent studies
\cite{PerezReche2003,PerezReche2004b}, have shown how the critical
point can be characterized in systems of moderate size. The key point
is to detect the so-called ``spanning'' avalanches which are the
magnetization jumps that involve a set of spins that spans the whole
finite system (e.g. cubic lattice) from one face to the opposite one.
By this method avalanches in finite systems can be classified as
non-spanning, 1D-spanning, 2D-spanning, or 3D-spanning.  The average
numbers $N_1$, $N_2$ of 1D and 2D-spanning avalanches display a peak
at a value of $\sigma$ that shifts with system size $L$ and tends to
$\sigma_c$ when $L \rightarrow \infty$.  The numerical data can then
be scaled according to the FSS hypothesis\cite{PerezReche2003}
\begin{equation}
N_{\alpha} = L^\theta {\tilde N}_{\alpha} (u L^{1/\nu})
\end{equation} 
where $\alpha=1,2$. The exponent $\nu=1.2\pm0.1 $ characterizes the
divergence of the correlation length ($\xi\sim
(\sigma-\sigma_c)^{-\nu}$) whereas $\theta=0.10\pm0.02$ characterizes
the divergence of the number of critical avalanches. The scaling
variable $u(\sigma)$ is analytic and measures the distance to the
critical point. It can be fitted by the second order expression
\begin{equation}
u (\sigma)=\frac{\sigma-\sigma_c}{\sigma_c} + 
A \left ( \frac{\sigma-\sigma_c}{\sigma_c}\right )^2
\label{upaco}
\end{equation}
with $\sigma_c=2.21$ and $A=-0.2$.  The behaviour of the 3D-spanning
avalanches is more complex because they are of two different kinds:
(i) critical 3D-spanning avalanches that behave like the 1D and 2D
ones and (ii) subcritical 3D-spanning avalanches that will corespond
to the $\Delta m$ discontinuity in the thermodynamic limit. The
analysis is more difficult and requires a double finite-size-scaling
technique.  This will not be used in the present paper. Instead we
will focus only on the behaviour of the average numbers $N_1$ and
$N_2$ in the presence of vacancies and propose a FSS hypothesis by
using a bivariate scaling variable $u(\sigma,c)$ that allows to study
the full $\sigma-c$ diagram.

In section \ref{Model}, we define the model and the dynamics.  In
section \ref{numerical}, we present results of the numerical
simulations. In section \ref{FSS}, we formulate the FSS hypothesis and
determine the critical line. In section \ref{Bivariate}, we propose
approximations to the bivariate scaling variable $u(\sigma,c)$. In
section \ref{Discussion}, we discuss the interplay between vacancies
and avalanches and finally, in section \ref{Conclusions}, we summarize
our main findings and conclude.

\section{Model and simulations}
\label{Model}

The diluted 3D-RFIM on a cubic lattice with $N$ sites ($N = L\times
L\times L$) is defined by the following Hamiltonian (magnetic
enthalpy):
\begin{equation}
{\cal H}= -\sum_{\langle ij\rangle}^{n.n.} c_i c_j S_i S_j -
\sum_{i=1}^N h_i c_i S_i -H\sum_{i=1}^N S_i c_i 
\end{equation}
where $S_i=\pm 1$ are Ising spin variables, $c_i=0,1$ indicates the
presence of a vacancy ($c_i=0$) or not ($c_i=1$) on each site, $h_i$
are quenched random fields gaussian distributed with zero mean and
standard deviation $\sigma$ and $H$ is the driving field. The first
sum extends over all distinct nearest-neighbour (n.n.) pairs.
Vacancies are quenched and randomly distributed over the lattice.
Their concentration is measured by $c=1-\sum_i c_i/N$.

The metastable dynamics is implemented as follows: the system is
externally driven by the field $H$ which is adiabatically swept from
$-\infty$ where the system is fully negatively magnetized ($S_i=-1$)
to $+\infty$. ($S_i=+1$). The spins flip according to a local
relaxation dynamical rule,
\begin{equation}
S_i = \sign(\sum_j S_j c_j + h_i + H)
\label{dynamics}
\end{equation} 
where the sum extends over all the n.n. of $S_i$. When a spin flips,
it may trigger an avalanche. The unstable spins are flipped
synchronously until a new stable situation is reached.

\begin{center}
\begin{figure}[th]
  \includegraphics[width=8cm,clip]{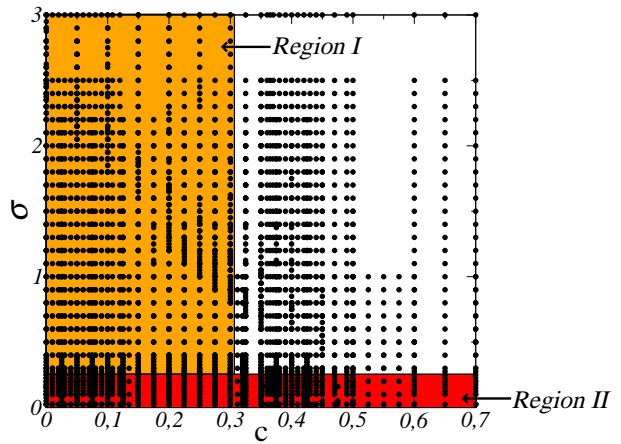}
\caption{\label{FIG1} 
  (Color online) Coordinates of the points studied by numerical
  simulations in the $\sigma-c$ diagram. The finite size scaling
  analysis presented in section \ref{FSS} is performed in regions I
  and II.}
\end{figure}
\end{center}

The hysteresis loop is obtained by computing the magnetization
\begin{equation}
m=\sum_{i=1}^N S_i c_i / N 
\end{equation}
as a function of the applied field $H$. Magnetization avalanches are
recorded along the whole increasing field branch and their spanning
properties are analyzed by using ``mask'' vectors (as explained in
Ref.~ \onlinecite{PerezReche2003}) that allow to classify them as
non-spanning, 1D-spanning, 2D-spanning and 3D-spanning. In this work
we shall mainly study the number of spanning avalanches of each kind
which are recorded in the full upwards branch.  These numbers, $N_1$,
$N_2$ and $N_3$ which depend on $L$, $\sigma$ and $c$ corresponds 
to averages over more than $10^4$ realizations with different 
random fields and random vacancy positions. The disorder averages 
are denoted by the symbol $\langle\cdot\rangle$.  
We study sytems of sizes ranging from $L=8$ to $L=64$
in a number of points on the $\sigma-c$ diagram, as indicated
schematically in Fig. \ref{FIG1}.
\section{Numerical results}
\label{numerical}
\begin{center}
\begin{figure}[th]
  \includegraphics[width=8cm,clip]{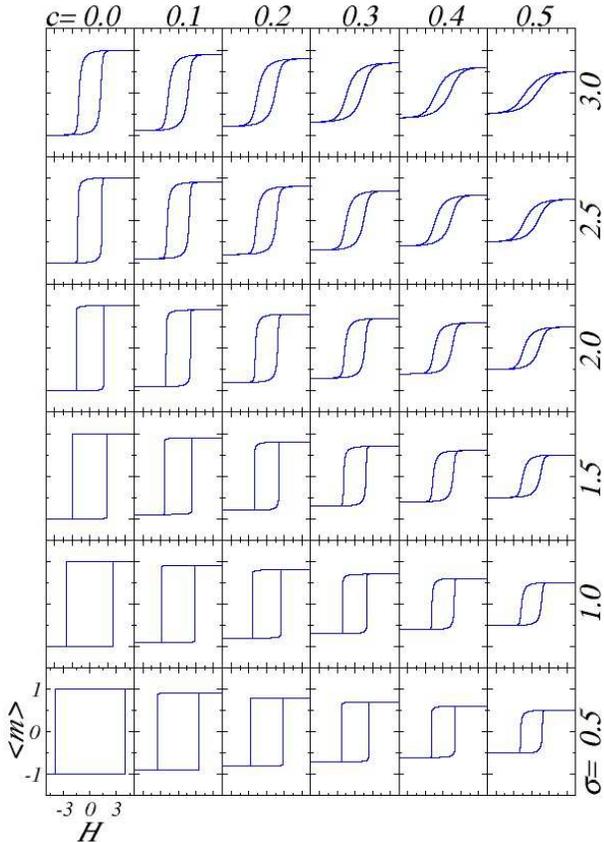}
\caption{\label{FIG2}   (Color online) 
  Average hysteresis loops corresponding to a system of size $L=32$ for
  different values of $\sigma$ and $c$ as indicated.}
\end{figure}
\end{center}

\begin{center}
\begin{figure}[th]
  \includegraphics[width=8cm,clip]{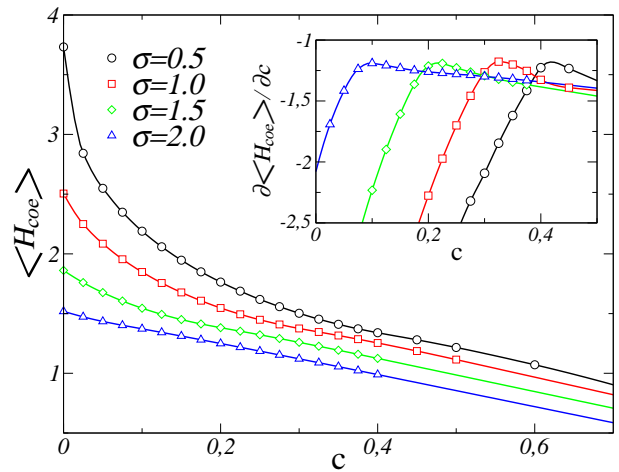}
\caption{\label{FIG3} 
  (Color online) Coercive field as a function of the vacancy
  concentration $c$ for different values of the amount of disorder
  $\sigma$. The inset shows the behaviour of the numerical derivative
  $\partial H_{coe}/\partial c$ which exhibits a maximum on the
  transition line. Data correspond to averages in a system of size
  $L=64$.}
\end{figure}
\end{center}

\begin{center}
\begin{figure}[th]
  \includegraphics[width=8cm,clip]{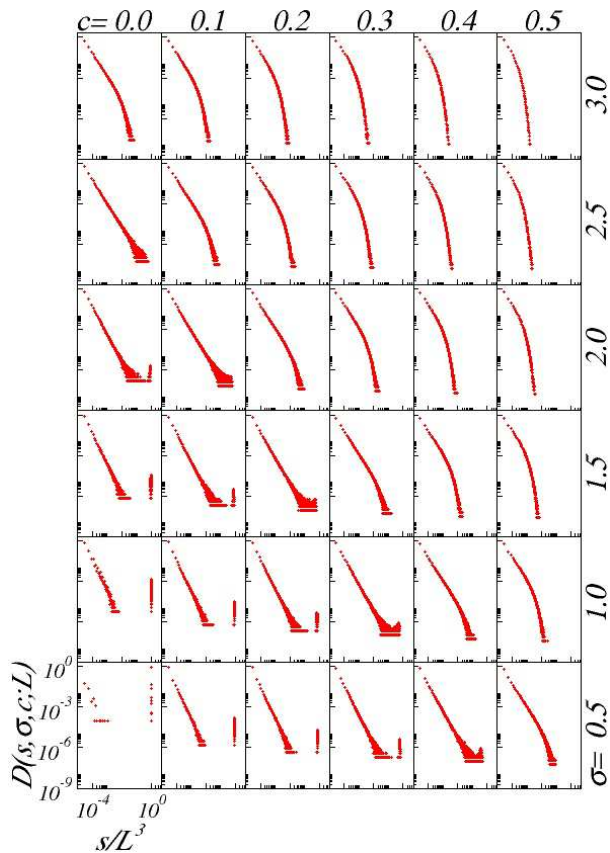}
\caption{\label{FIG4} 
  (Color online) Avalanche size distributions corresponding to a
  system with $L=32$ at different values of $c$ and $\sigma$ as
  indicated. Data are represented in log-log scale.}
\end{figure}
\end{center}

\begin{center}
\begin{figure}[th]
  \includegraphics[width=8cm,clip]{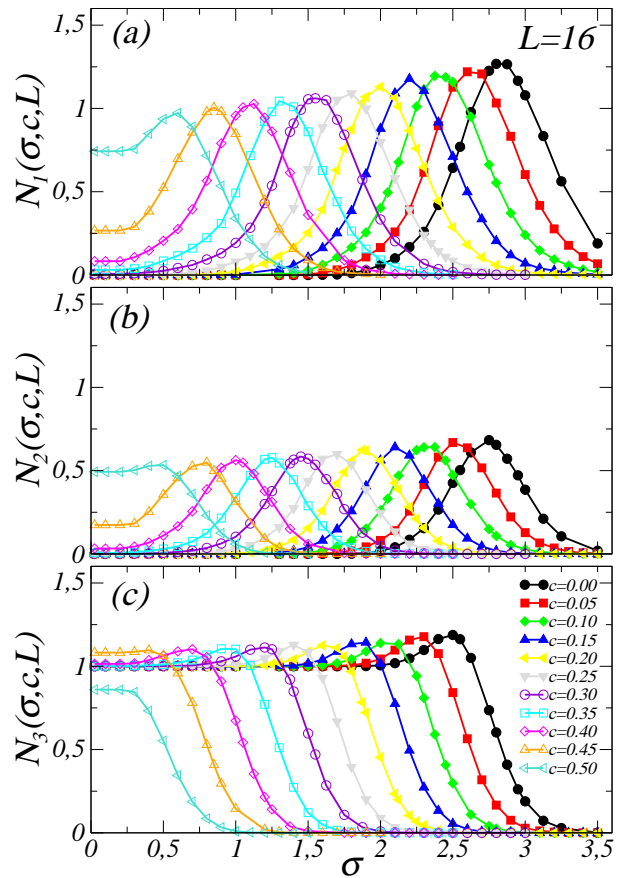}
\caption{\label{FIG5} 
  (Color online) Average number of 1D-spanning avalanches (a),
  2D-spanning avalanches (b) and 3D-spanning avalanches (c) as a
  function of $\sigma$ for different values of the vacancy
  concentration $c$, as indicated by the legend. Lines are guides to
  the eye. Data corresponds to numerical simulations of a system with
  size $L=16$. }
\end{figure}
\end{center}

\begin{center}
\begin{figure}[th]
  \includegraphics[width=8cm,clip]{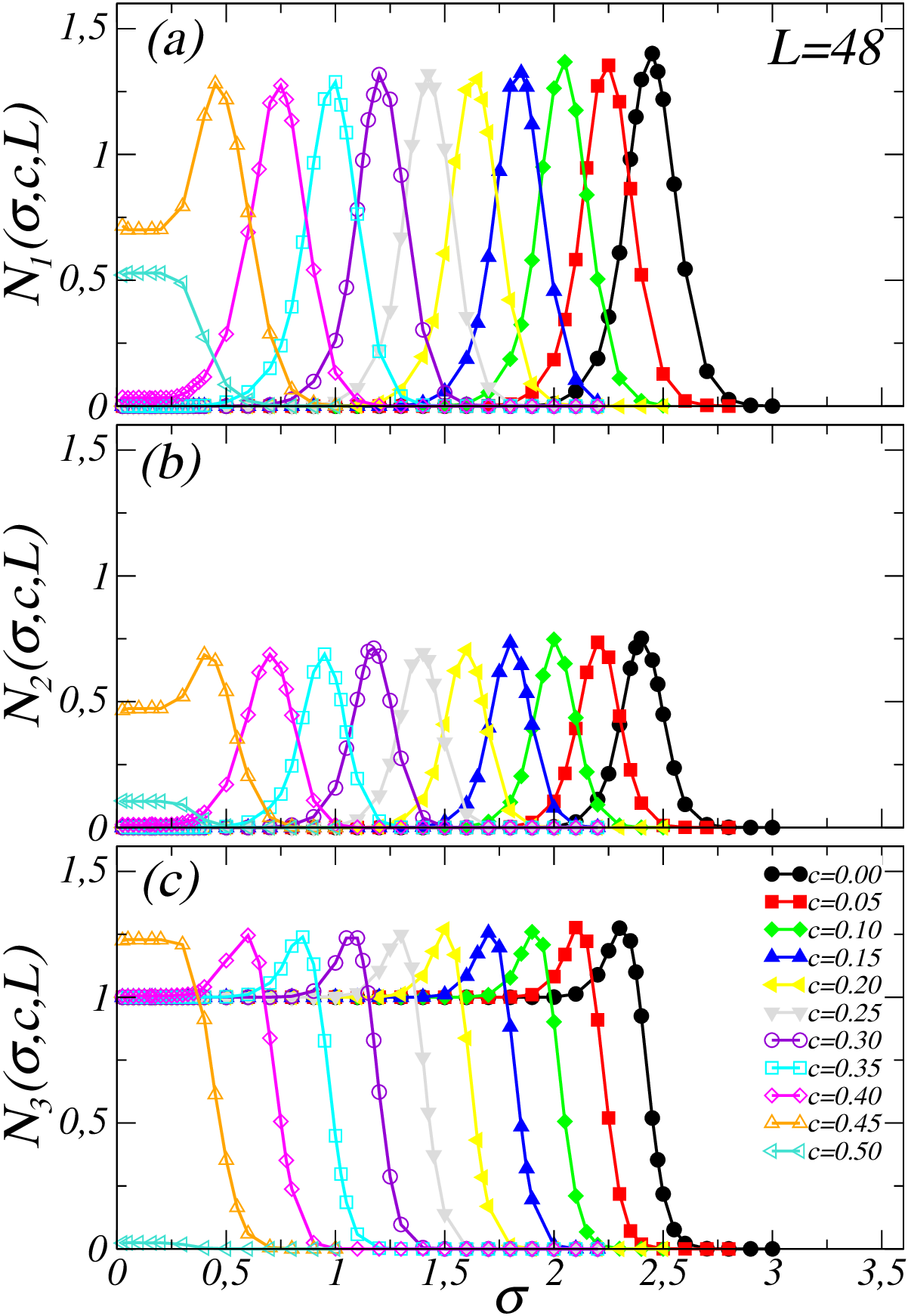}
\caption{\label{FIG6}  
  (Color online) Average number of 1D-spanning avalanches (a),
  2D-spanning avalanches (b) and 3D-spanning avalanches (c) as a
  function of $\sigma$ for different values of the vacancy
  concentration $c$, as indicated by the legend. Lines are guides to
  the eye. Data corresponds to numerical simulations of a system with
  size $L=48$.}
\end{figure}
\end{center}

The general evolution of the average hysteresis loops as a function of
$\sigma$ and $c$ is shown in Fig.~\ref{FIG2}. One can observe the
transition from discontinuous loops to smooth loops when $\sigma$ or
$c$ are increased. It can also be seen that the saturation
magnetization decreases with increasing $c$.
Figure \ref{FIG3} shows the behaviour of the coercive field $\langle
H_{coe} \rangle$ as a function of the concentration of vacancies for
different values of $\sigma$. As can be seen, $\langle H_{coe} \rangle$ 
decreases with increasing $c$ and increasing $\sigma$.  The
behaviour with increasing $c$ exhibits an inflection point at the
transition as can be seen in the inset of Fig.~\ref{FIG3} which shows
the numerical derivative of the $\langle H_{coe} \rangle$ with respect
to $c$. Such an inflection point does not exist in the non-diluted
model when the coercive field is plotted as a function of $\sigma$.
This feature, which can be of interest for the determination of the
critical point in experiments, is probably related to the fact that
$\langle H_{coe} \rangle$, as a function of $c$ should vanish at $c \leq 1$ whereas as a
function of $\sigma$ it only vanishes assymptotically when $\sigma
\rightarrow \infty$.
Fig.~\ref{FIG4} shows the distribution $D(s; \sigma,c,L)$ of avalanche
sizes (the size $s$ of an avalanche is the number of spins flipped)
for the same cases as in Fig.~\ref{FIG2} in log-log scale.  The
histograms include all avalanches irrespective of their spanning
properties. The qualitative picture is that power law distributions
are obtained along a critical line with an exponent that seems to be
the same for all values of $c$.  Apparently, no differences can be
observed when comparing the transition induced by changing $\sigma$
from the transition induced by changing $c$.  Below the critical line
the distributions show a peak for large values of $s$ which correspond
to the 1D, 2D and 3D spanning avalanches.  Above, the distributions
have an exponentially damped character.
Figure \ref{FIG5} shows the average number of 1D, 2D and 3D spanning
avalanches as a function of $\sigma$ for increasing values of the
vacancy concentration $c$ ranging from $0$ to $0.5$. Data corresponds to a system with size $L=16$.
The same information is displayed in Fig. \ref{FIG6} for a system with
size $L=48$.
The behaviour for small and intermediate vacancy concentration is
qualitively similar to that found for the non-diluted model
\cite{PerezReche2003}. The average numbers $N_1$ and $N_2$ display
peaks, whereas $N_3$ shows a peak on the edge of a step function. Note
that for $L=16$ the peak height in $N_1(\sigma,c,L)$ and
$N_2(\sigma,c,L)$ seem to decrease with increasing $c$.  This
behaviour, however, is much less apparent for larger systems ($L=48$).
Thus, it is possibly due to a finite size effect.

\begin{center}
\begin{figure}[th]
  \includegraphics[width=10cm,clip]{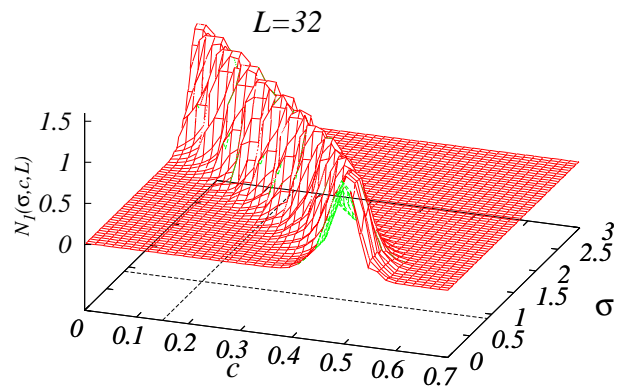}
\caption{\label{FIG7}  (Color online) Surface plot representing $N_1(\sigma, c,L)$ for $L=32$. The dashed lines on the basal plane represent the position of the cuts in Fig.\ref{FIG8} at $c=0.15$ and $\sigma=0.9$.}
\end{figure}
\end{center}

At higher concentrations ($c>0.4$) $N_1$ and $N_2$ begin to develop a
flat plateau at low $\sigma$. The reason for this plateau can be well
understood by looking at the 3d-plot in Fig. \ref{FIG7}, which
represents the average number $N_1(\sigma,c,L)$ for $L=32$. The
plateau in the constant $c$ cuts of Figs. \ref{FIG5} and \ref{FIG6} is
due to the fact that the crest of the $N_1$ and $N_2$ functions bends
towards the $\sigma=0$ axis.
\section{Finite Size Scaling hypothesis}
\label{FSS}
\begin{center}
\begin{figure}[th]
  \includegraphics[width=8cm,clip]{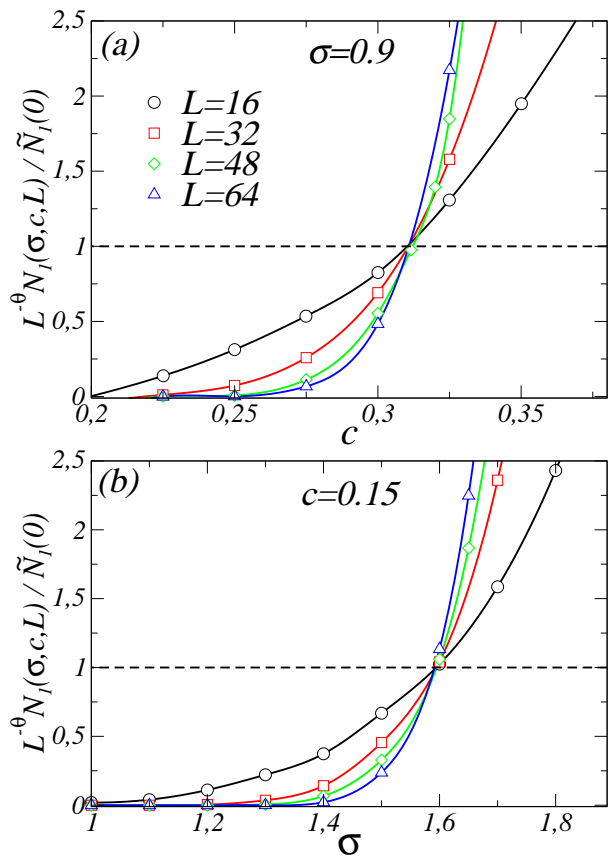}
\caption{\label{FIG8} 
  (Color online) Examples of crossing points on the critical line on
  cuts (a) parallel to the $c$ axis and (b) parallel to the $\sigma$
  axis. The different symbols correspond to different system sizes as
  indicated by the legend. Continuous lines are guides to the eye.
  The horizontal dashed-line indicates the height $1$ where the curves
  are supposed to cross according to Eq. \ref{FSSu0}.  }
\end{figure}
\end{center}

\begin{center}
\begin{figure}[th]
  \includegraphics[width=8cm,clip]{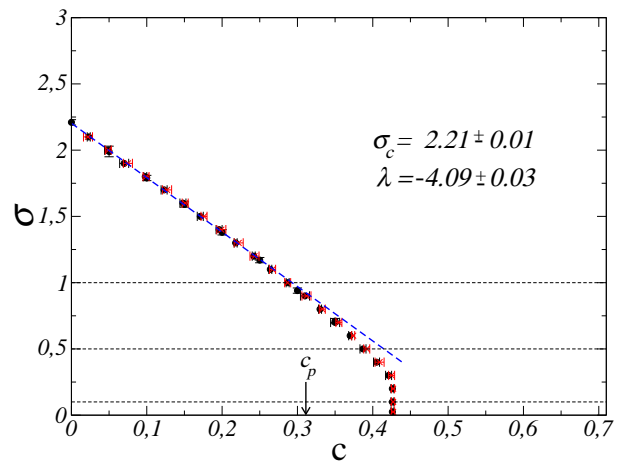}
\caption{\label{FIG9}  
(Color online) Critical line in the $\sigma-c$ diagram 
  determined from the crossing points in $N_1$  ($\bullet$) and $N_2$
  ($\times$).  The dashed line is the fit discussed in the text, and
  the thin discontinuous lines indicate the cuts along which the
  correlation in Fig.~\ref{FIG14} is computed.}
\end{figure}
\end{center}

The hypothesis that we want to check numerically is that, in the
presence of vacancies, the critical point found at $c=0$ transforms into a
critical line for a wide range of concentrations. Thus the critical
exponents found previously should be equally valid for the description
of the behaviour of the average numbers $N_1$ and $N_2$ with $c>0$.
According to this hypothesis we shall propose the following
corresponding FSS behaviour:
\begin{equation}
N_{\alpha}(\sigma, c, L) = L^\theta {\tilde N}_{\alpha} (u L^{1/\nu})
\label{FSS2}
\end{equation}
where $\alpha=1,2$ and $u(\sigma,c)$ is a bivariate scaling variable
measuring the distance to the critical line. The exponents $\theta$
and $\nu$ as well as the functions $\tilde N_{\alpha}$ were already
found in previous works \cite{PerezReche2003}. Therefore, the
hypothesis is quite strong and indicates that all the $N_1$ and $N_2$
data corresponding to different sizes $L$, different vacancy
concentrations $c$ and different amounts of disorder $\sigma$ must
collapse into an already known function. The only freedom is in the
determination of the bivariate scaling variable $u$ that should be
analytic. Before constructing it in the next section, we can make a
first test of Eq.~\ref{FSS2}, by cheking the scaling on the critical
line. Note that, setting $u=0$, Eq. \ref{FSS2} becomes:
\begin{equation}
N_{\alpha}(\sigma, c, L) = L^\theta {\tilde N}_{\alpha} (0)
\label{FSSu0}
\end{equation}
where $\sigma$ and $c$ should be on the critical line. Since we know
(from Ref. \onlinecite{PerezReche2003}) that $\theta=0.10$, ${\tilde
  N}_1 (0)=0.12 $ and ${\tilde N}_2 (0)=0.07 $, we can deduce that the
different curves $N_{\alpha}(\sigma, c, L)/L^\theta {\tilde
  N}_{\alpha} (0)$ should cross at height $1$ on the critical line,
independently of $L$.  Two examples are shown in Fig. \ref{FIG8} that corresponds to two cuts (one at constant
$\sigma$ and the other at constant $c$) in the $\sigma-c$ diagram. As
can be seen, the critical line can be determined with high accuracy.
By analyzing a large number of such $\sigma$ and $c$ cuts we have
constructed it. The result is shown in Fig.  \ref{FIG9}. Note that the
process can be repeated independently with $N_1$ and $N_2$. The two
independent lines overlap almost perfectly.

The obtained critical line is linear up to $c\simeq 0.3$. A least
squares fit gives $\sigma_c(c) = \sigma_c(0)+ \lambda c$ with
$\sigma_c(0) = 2.21\pm 0.01$ and $\lambda= -4.09\pm 0.03 $. The value
$\sigma_c(0)= 2.21$ is in total agreement with the previous estimate
for the non-diluted model \cite{PerezReche2003}.

It is remarkable that the finite size scaling hypothesis allows the
collapse of the data up to large values of $c$, far from the point $c=0$ where
the scaling function and the exponents where determined. It
is also remarkable that scaling works even after the bend that is
observed for $c>0.3$. (Note that the crossing point shown in Fig.
\ref{FIG8}(a) corresponds to a value of $\sigma$ where the critical
line is not linear.)

For small values of $\sigma$, the critical line displays a vertical behaviour.  
The critical value of the vacancy concentration
$c_c$ above which the hysteresis loops do not display a
discontinuity can be fitted to $c_c=0.426 \pm 0.003$

\section{Bivariate scaling variable}
\label{Bivariate}
In general the bivariate scaling variable is a function that can be
expanded as:
\begin{equation}
u(\sigma,c) = a_0 + a_1 \sigma + a_2 c + a_3 \sigma c + a_4 \sigma^2 + a_5 c^2 + \dots
\label{uexpansion}
\end{equation}
Since $c$ and $\sigma$ are not necessarily very
small along the critical line, it is difficult to know a priori how many terms in
the expansion will be needed in order to find a good scaling collapse.
The direct determination of a large number of coefficients from the
numerical data is difficult. Therefore we shall take a different
strategy taking into account, as much as possible, the previously
known data.

\begin{center}
\begin{figure}[th]
  \includegraphics[width=8cm,clip]{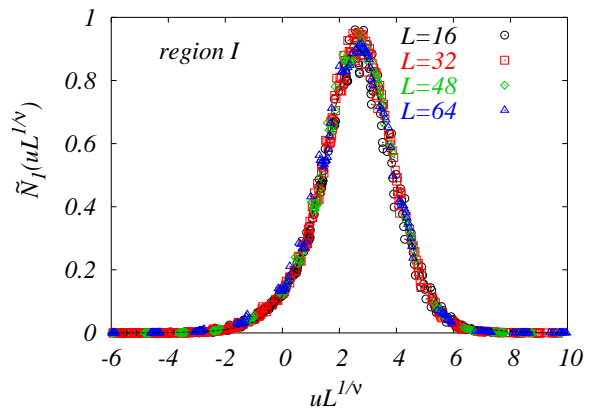}
\caption{\label{FIG10} (Color online) Finite-size-scaling collapse of the average number of 1D-spanning 
  avalanches in region I. The continuous line shows the Lorentzian
  function in Eq.~\ref{scalfunc1}.}
\end{figure}
\end{center}

\begin{center}
\begin{figure}[th]
  \includegraphics[width=8cm,clip]{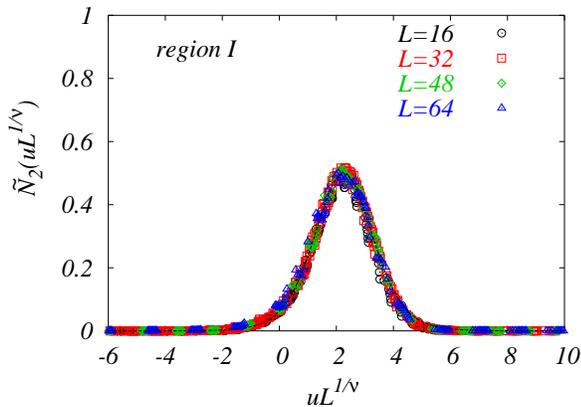}
\caption{\label{FIG11} (Color online) Finite-size-scaling collapse of the average number of 2D-spanning avalanches in region I. The continuous line shows the Lorentzian function in Eq.~\ref{scalfunc2}.}
\end{figure}
\end{center}

\begin{center}
\begin{figure}[th]
  \includegraphics[width=8cm,clip]{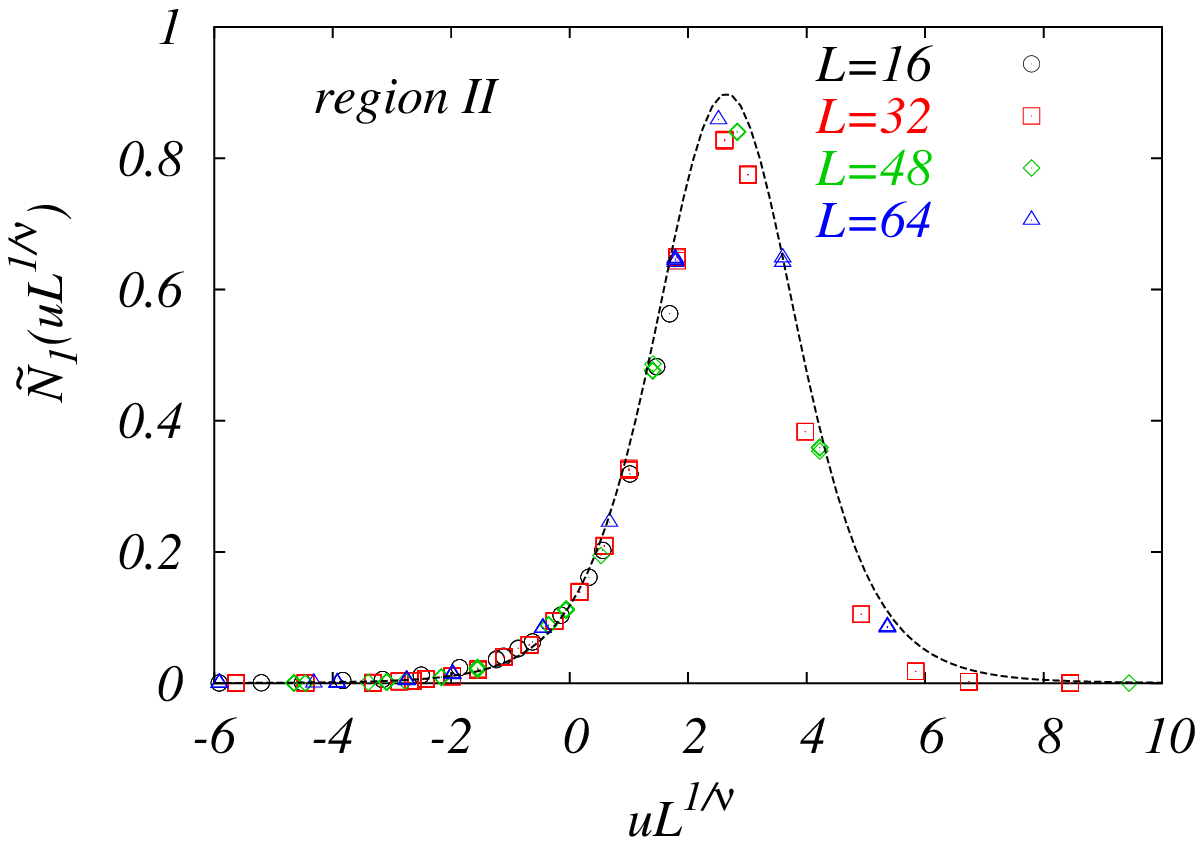}
\caption{\label{FIG12}  (Color online) Finite size scaling collapse of the average number of 1D-spanning avalanches in region II. The continuous line shows the lorentzian funcion in Eq.~\ref{scalfunc1}.}
\end{figure}
\end{center}

\begin{center}
\begin{figure}[th]
  \includegraphics[width=8cm,clip]{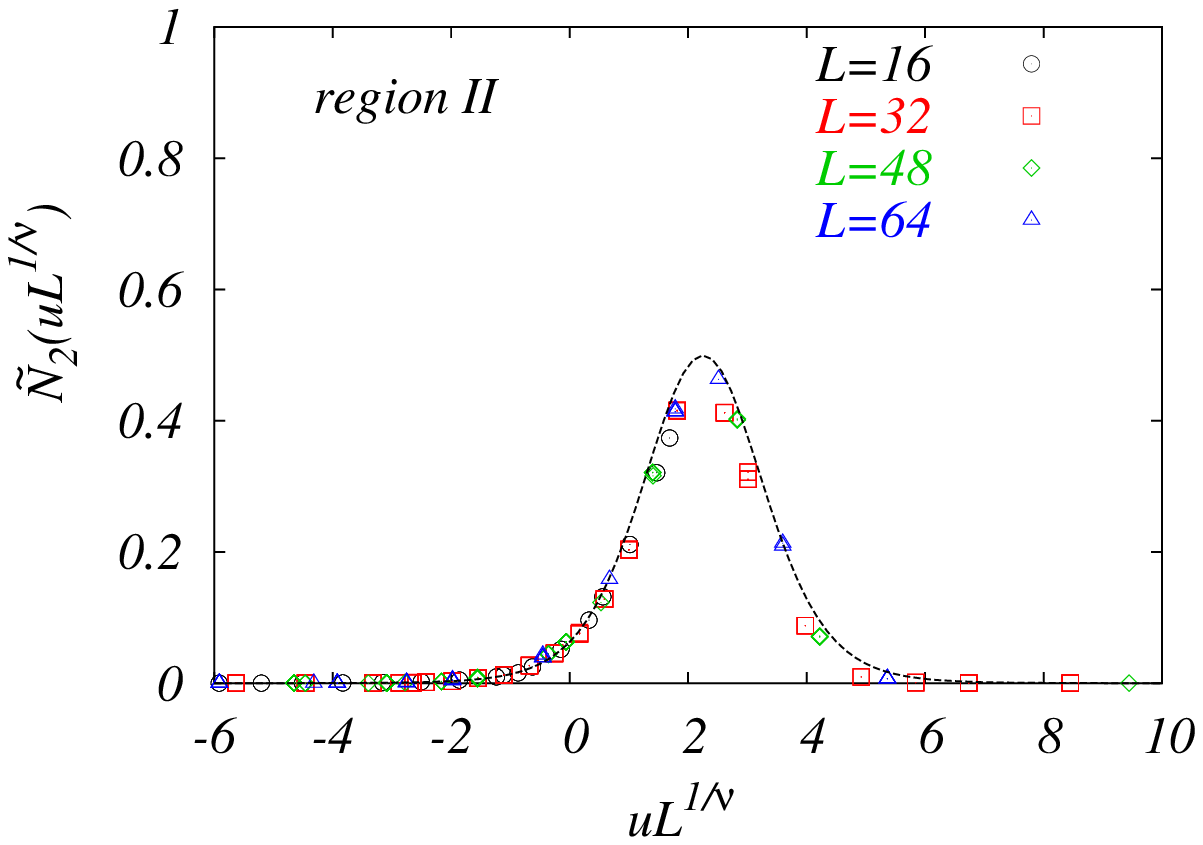}
\caption{\label{FIG13}  (Color online) Finite size scaling collapse of the average number of 2D-spanning avalanches in region II. The continuous line shows the lorentzian funcion in Eq.~\ref{scalfunc2}.}
\end{figure}
\end{center}

As a first step we will concentrate in the region $c\leq 0.3$ where
the coexistence line shows a linear behaviour and we will try to use
an expansion up to quadratic terms only. By forcing that the condition
$u=0$ is satisfied on the fitted coexistence line, we deduce that $u$
satisfies:
\begin{equation}
\label{scalv1}
u(\sigma,c) = (\sigma - \sigma_c - \lambda c)(b_0 + b_1 \sigma + b_2 c)
\end{equation}
We should also consider the fact that the scaling variable is known to
be well described by a second order expansion (up to $\sigma^2$) for
$c=0$ as indicated in Eq. \ref{upaco}.  After some algebra one can
determine the two parameters $b_0$ and $b_1$:
\begin{eqnarray}
b_0&=& (1-A)/\sigma_c = 0.543 \pm 0.002\\
b_1&=& A /\sigma_c^2 = -0.041 \pm 0.001
\end{eqnarray}
Therefore we are left with a single free parameter $b_2$ that should
allow to collapse into a single curve all the data in the scaling
region, for different values of $\sigma$, $c$ and $L$. We have
considered all the available data in region I of Fig. \ref{FIG1}. Note
also that the same $b_2$ parameter must be used to scale both $N_1$
and $N_2$ data. The best two collapses are shown in Fig. \ref{FIG10}
and Fig. \ref{FIG11} for $b_2=-0.13$.
Note that the data for $c=0$ are also included in this plot. Therefore
we are obtaining two scaling functions $\tilde N_1$ and $\tilde N_2$
compatible with those of Ref.~\onlinecite{PerezReche2003}. In that
reference the scaling functions were approximated by Gaussians,
although it was also shown that there were systematic deviations. In
this work we have tried to fit the data with more complex functions
(with three free parameters). We have found a very good $\chi^2$ by
using the following modified Lorentzians, which are represented by a
continuous line on top of the data in Figs.~\ref{FIG10} and
\ref{FIG11}.
\begin{equation}
\tilde N_1 (x) =\frac{1}{(1.73-0.53x+0.10x^2)^{3.9}}
\label{scalfunc1} 
\end{equation}
\begin{equation}
\tilde N_2 (x) =\frac{1}{(1.83-0.59x+0.13x^2)^{4.6}}
\label{scalfunc2}
\end{equation}
As a second step we try to build $u(\sigma,c)$ for the
data very close to the $\sigma=0$ axis. In this region II (see Fig.
\ref{FIG9}) the transition line is again quite linear and in fact is
almost vertical. This means that to measure the distance to the
critical line it should be sufficient to use the variable $(c-c_c)$. We have
considered the following second-order expansion:
\begin{equation}
\label{scalv2}
\frac{u(c)}{k'}=\frac{c-c_c}{c_c} + B \left ( \frac{c-c_c}{c_c} \right ) ^2
\end{equation}
Note that $k'$ is not a free parameter. It can be fixed by imposing
that the definitions of the scaling variables (\ref{scalv1}) and (\ref{scalv2}) coincide at $\sigma=0$
and $c=0$. Thus $k'=(A-1)/(B-1)$. The only free parameter for the
collapse of the data is $B$.  Best collapses are shown in Fig.
\ref{FIG12} and \ref{FIG13} for $N_1$ and $N_2$ respectivelly using
the best choice $B=-0.2$ (thus $k'=1$.)
The continuous lines in both figures correspond to the same lines as
in Figs.~\ref{FIG10} and \ref{FIG11}. We can thus conclude that we
have built two good approximations (given by Eq. \ref{scalv1} and
\ref{scalv2}) in regions I and II, to the unique scaling variable
which will display a more complex behaviour in the intermediate region
were the critical line bends, probably with higher order terms.

\section{Discussion}
\label{Discussion}
In this section we try to understand why the critical line exhibits
such a curvature. There must be a physical reason that goes beyond the
mere effect of the dilution of the system and unstabilizes even more
the phase with the ferromagnetic-like discontinuity. We propose that
the effect is related to the percolation of vacancies above
$c_p=0.3116$, a value which is, indeed, very close to the limit where
the critical line loses its linearity.  To justify this hypothesis
numerically we have studied, for each particular realization of
disorder the distribution of the clusters of vacancies and the
position of the avalanches. In particular we have determined the
spatial position of the largest vacancy cluster (which above $c_p$
will correspond to the percolating cluster in the thermodynamic
limit). It is clear that the neigbouring sites of this percolating cluster of vacancies are an easy path for the propagation of an avalanche
since these sites have a smaller number of neighbours. To
distinghuish such sites we have defined a local flag that takes values
$b_i=1$ when a site belongs to the border of the largest cluster of vacancies or $b_i=0$ otherwise.  We have also recorded the largest
avalanche during the $H$-scan (which will
correspond to the spanning avalanche below the critical line in the thermodynamic limit) and
marked its position with a flag $\epsilon_i=1$.  With these two
variables we have defined the correlation between the border of the
largest cluster of vacancies and the largest avalanche as:
\begin{equation}
\rho_{\epsilon,b}= \frac{\langle \frac{1}{N} \sum \epsilon_i b_i \rangle - 
\langle \frac{1}{N} \sum \epsilon_i \rangle  
\langle \frac{1}{N} \sum_i b_i \rangle }
{\sqrt{\langle \frac{1}{N} \sum \epsilon_i^2 \rangle 
- \langle \frac{1}{N} \sum \epsilon_i \rangle^2 }    
\sqrt{\langle \frac{1}{N} \sum b_i^2 \rangle - 
\langle \frac{1}{N} \sum b_i \rangle^2}}
\end{equation}
\begin{center}
\begin{figure}[th]
  \includegraphics[width=8cm,clip,]{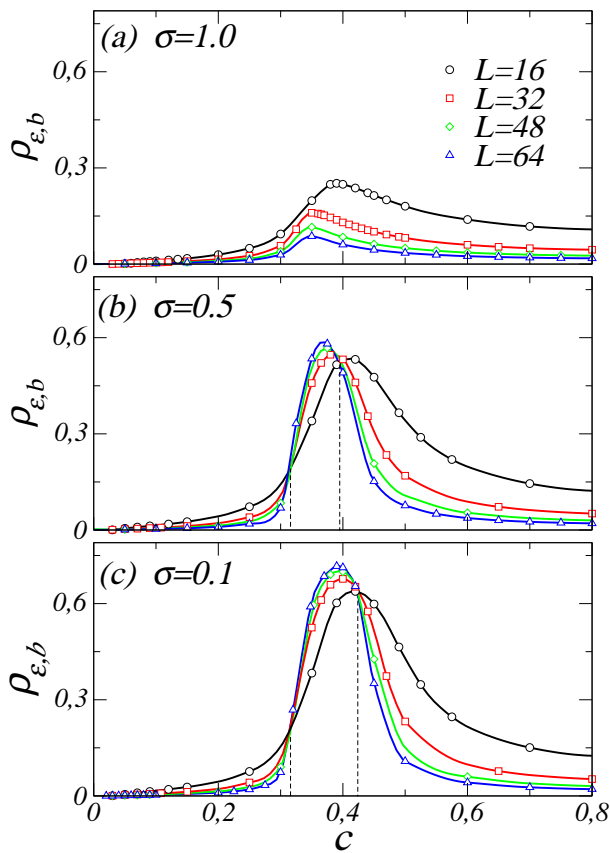}
\caption{\label{FIG14} 
  (Color online) Correlation between the border of the largest cluster
  of vacancies and the largest avalanche as a function of $c$. Data
  corresponding to different system sizes are represented by different
  symbols as indicated by the legend. The curves correspond to cuts in
  the phase diagram at $\sigma=1$ (a), $\sigma=0.5$ (b) and
  $\sigma=0.1$ (c).  }
\end{figure}
\end{center}
Note that since $\epsilon_i$ and $b_i$ take values $1,0$ only, the
power $2$ in the first bracket inside the square roots can be
suppressed.  This correlation is equal to $1$ when the spanning
avalanche sits exactly on the border of the spanning cluster of vacancies.
The behaviour of $\rho_{\epsilon,b}$ as a function of $c$ is shown in
Fig.~\ref{FIG14} for three different values of $\sigma$ that
correspond to the dashed lines indicated in Fig.~\ref{FIG9} and for
increasing system sizes as indicated by the legend. The important
observation is that the curves for $\sigma=0.5$ and $\sigma=0.1$
exhibit two crossing points. One is located at $c_p$ and the other on
the critical line (it thus shifts with $\sigma$). For a concentration
of vacancies below $c_p$ or above the critical line, the behaviour of
the curves with increasing $L$ indicates that the correlation vanishes
in the thermodynamic limit, whereas in the region between the two
crossing points the correlation increases with increasing system size.
A value $\rho=1$ is probably not reached since the spanning avalanche
is larger than the border of the percolating cluster of vacancies.
By this analysis, we have thus identified the origin of the curvature
of the critical line: when vacancies percolate the spanning avalanche
propagates along the border of the percolating cluster of vacancies.
The propagation in such a constrained environment decreases the 
amount of disorder needed to break the infinite macroscopic avalanche 
into small microscopic jumps.  However, as shown in the previous 
section this mechanism does not changes the values of the critical exponents.

\section{Summary  and conclusions}
\label{Conclusions}
We have analyzed the influence of dilution on the critical properties
of the 3D-RFIM at $T=0$ with metastable dynamics. We have shown that
the critical point associated to the change in the shape of the
hysteresis loop from discontinuous to continuous loops becomes a
critical line which we have located in the $\sigma-c$ phase diagram.
The critical properties close to this line are characterized by the
same critical exponents as in the non-diluted model. This result
indicates that it should be possible to find RG arguments showing that
there is a unique fixed point at $T=0$ in the disorder parameter
space, that includes, at least, both random fields and dilution
\cite{Sethna2005}. We have computed quadratic approximations to the scaling
variable in two different zones of the phase diagram that allow for a
bivariate finite-size-scaling collapse on a universal scaling
function. Finally we have proposed an explanation for the the
curvature observed in the critical line when the concentration of
vacancies increases above the percolation limit: the spanning
avalanche that is responsible for the discontinuity of the hysteresis
loops, has a tendency to follow the neigbourhood of the percolating
cluster of vacancies.

\section{Acknowledgements}

We acknowledge fruitful discussions with M.L.Rosinberg,
F.J.P\'{e}rez-Reche and A.Planes.  This work has received financial
support from projects MAT2004-01291 (CICyT, Spain) and SGR-2001-00066
(Generalitat de Catalunya).  X.I. acknowledges a grant from DGI-MEC
(Spain).

\end{document}